\documentclass[
twocolumn,
pra,showpacs,preprintnumbers,amsmath,amssymb]{revtex4}

\usepackage{graphicx}
\usepackage{amssymb}
\usepackage{amsmath}
\usepackage{dsfont}
\usepackage{bm}
\usepackage{mathrsfs}

\begin{document}

%%%%%%%%%%%%%%%%%%%%%%%%%%%%%%%%%%%%%%%%%%%%%%%%%%%%%%%%%%%%%%%%%%%%%%%%%%%%%%%%%%%

\title{Perfect imaging with positive refraction in three dimensions}
\author{Ulf Leonhardt and Thomas G. Philbin}
\affiliation{
University of St Andrews, School of Physics and Astronomy,
North Haugh, St Andrews KY16 9SS, UK
}
\date{\today}
\begin{abstract}
Maxwell's fish eye has been known to be a perfect lens within the validity range of ray optics since 1854. Solving Maxwell's equations we show that the fish--eye lens in three dimensions has unlimited resolution for electromagnetic waves. 
\end{abstract}
\maketitle

Perfect imaging with positive--index materials has been discussed since 1854 \cite{Maxwell}, but only for light rays and not for waves \cite{Luneburg,BornWolf}. In fact, the defining property of such ideal optical instruments \cite{BornWolf} is the perfect focussing of rays: for each point in a region of space, all emitted rays meet in a corresponding point in the image space. Yet it is the wave nature of light that sets the resolution of optical instruments. Only very recently \cite{LeoPerfect}, one of us proved that the archetype of all ideal optical instruments \cite{Luneburg}, Maxwell's fish eye \cite{Maxwell} perfectly images light waves in two dimensions. 
Here we show that the fish eye in three dimensions has infinite resolution as well, provided its medium is impedance--matched. Absorption does not appear to significantly reduce the image quality, in contrast to imaging with negative refraction \cite{Pendry,Stockman}. However, fish--eye lenses contain both the source and the image inside the medium and impedance--matched devices are still difficult to make in practice. On the other hand, our case supports the idea \cite{LeoPerfect} that perfect imaging is not necessarily caused by the amplification of evanescent waves \cite{Pendry} but rather by the geometry of light \cite{LeoPhilbin,Schleich,Note}. Such conceptual insights may be vital for the future direction of technical developments for perfect imaging.

According to Fermat's principle \cite{BornWolf,LeoPhilbin}, light rays follow extremal optical paths with a path length measured by the refractive index $n$ --- geodesics. Optical materials thus establish virtual spatial geometries for rays \cite{LeoPhilbin,Genov}. Whether these geometries are also valid for waves depends on the type of wave propagation. In two--dimensional structures, typical of integrated optics, the polarization of light decides whether the structures are perceived as geometries: for materials with purely electrical response TE polarization \cite{LL8} is required \cite{LeoPerfect}. In three dimensions, the material must be impedance--matched for establishing a virtual geometry for electromagnetic fields \cite{LeoPhilbin}, with equal electric permittivity $\varepsilon$ and magnetic permeability $\mu$,
%%%%%%%%%%%%%%%%
\begin{equation}
\varepsilon = \mu = n \,.
\end{equation}
%%%%%%%%%%%%%%%%
This crucial condition has not been considered in the previous treatment of Maxwell's fish eye with Maxwell's equations \cite{Tai,Rosu}. Scalar waves \cite{MG} should not obey the Helmholtz or Schr\"odinger equation either, for perfect imaging in three dimensions, but rather a wave equation we consider here as well, Eq.\ (\ref{scalar}). In the previous wave theories of Maxwell's fish eye \cite{Tai,Rosu,MG,Guenneau}, perfect imaging was impossible. If, however, the medium or polarization is chosen such that the geometry of light is not restricted to rays but extends to waves, waves may be as perfectly imaged as rays \cite{LeoPerfect}.

Maxwell's fish eye turns out \cite{Luneburg} to establish a non--Euclidean geometry \cite{LeoTyc}, the 3D surface of the 4D hypersphere, by the refractive--index profile
%%%%%%%%%%%%%%%%
\begin{equation}
n = \frac{2}{1+r^2}
\label{fish}
\end{equation}
%%%%%%%%%%%%%%%%
where $r$ denotes the distance from the center measured in the characteristic length scale of the device. As the index profile (\ref{fish}) is radially symmetric, the trajectory of a light ray lies in a plane, due to the conservation of angular momentum. So, for light rays, the propagation in 3D is the same as in 2D where Maxwell's fish eye corresponds to the surface of a 3D sphere.
Here all the rays emitted from one point travel along the great circles, meeting again at the antipodal point, which proves that Maxwell's fish eye fits the definition of an ideal optical instrument \cite{BornWolf}. In physical space, the antipodal image of a point $\mathbf{r_0}$ turns out \cite{BornWolf} to appear at 
%%%%%%%%%%%%%%%%
\begin{equation}
\mathbf{r'_0} = - \frac{\mathbf{r_0}}{r_0^2}   \,.
\label{image}
\end{equation}
%%%%%%%%%%%%%%%%
In this paper we show that electromagnetic waves are perfectly imaged at $\mathbf{r'_0}$, apart from a phase delay of
%%%%%%%%%%%%%%%%
\begin{equation}
\varphi = \pi k
\label{phase}
\end{equation}
%%%%%%%%%%%%%%%%
where $k$ denotes the wavenumber in units of the inverse length scale of the fish eye. As the phase (\ref{phase}) is uniform, objects are not only faithfully but also coherently imaged. Furthermore, the phase (\ref{phase}) is linear in wavenumber and so in frequency if the refractive index is not frequency-dependent. In this case, the time delay between source and image is uniform as well. Note that the phase delay is different in 2D \cite{LeoPerfect}: $\pi \nu$ with $\nu^2 = k(k+1)$. In the following we prove perfect imaging and obtain the quantitative results (\ref{image}) and (\ref{phase}) by analyzing the electromagnetic Green's function.

{\it Electric and magnetic fields.---}  The Green's function $G$ describes the electric field of a stationary wave with wavenumber $k$ emitted by an elementary dipole at position $\mathbf{r_0}$ that may point in all three directions. The Green's function is a matrix also known, in electrical engineering, as a dyade or a bi--tensor with the first index referring to the electric--field strength at the spectator point $\mathbf{r}$ and the second describing the direction of the dipole source at $\mathbf{r_0}$. If not otherwise stated we use Cartesian coordinates. As any source is a weighted and directed distribution of elementary point dipoles, it is sufficient to consider the Green's function $G (\mathbf{r}, \mathbf{r_0})$ for perfect imaging. The Green's function obeys the wave equation 
%%%%%%%%%%%%%%%%
\begin{equation}
\nabla \times \frac{1}{n} \nabla \times G - n \, k^2 G = \iota \mathds{1} 
\label{green}
\end{equation}
%%%%%%%%%%%%%%%%
with the infinitely localized current density $\iota$.
In a stationary regime, the wave emitted at the source must disappear somewhere. Usually the wave ultimately disappears at $\infty$, but in perfect imaging the entire energy of the emitted wave is focused at the image and does not reach $\infty$ at all. In practice, the light will be absorbed at the image, for example in a lithographic photoresist or a detector. In any case, in order to maintain a stationary regime we must supplement the source by a drain at the image (\ref{image}) with phase delay (\ref{phase}) such that \cite{LeoPerfect}
%%%%%%%%%%%%%%%%
\begin{equation}
\iota = \delta (\mathbf{r}-\mathbf{r_0}) - \mbox{e}^{\mathrm{i}\varphi} \delta (\mathbf{r}-\mathbf{r'_0})    \,.
\label{iota}
\end{equation}
%%%%%%%%%%%%%%%%
The minus indicates that the drain is a source run in reverse. Finally, causality implies \cite{LeoPerfect} that $G$ is analytic in $k$ on the upper half complex plane and vanishing for $k\rightarrow\infty$ there. If $G$ meets all these requirements our case is settled: Maxwell's fish eye makes a perfect lens in 3D.

In order to deduce the Green's function it is advantageous to represent $G$ in terms of its magnetic field, the tensor $H$, as 
%%%%%%%%%%%%%%%%
\begin{equation}
G = \frac{\nabla \times H - \iota \mathds{1}}{n \, k^2}
\label{grep}
\end{equation}
%%%%%%%%%%%%%%%%
where we require $H$ to obey the wave equation 
%%%%%%%%%%%%%%%%
\begin{equation}
\nabla \times \frac{1}{n} \nabla \times H - n \, k^2 H = \nabla \times \frac{\iota \mathds{1}}{n}   \,.
\label{hwave}
\end{equation}
%%%%%%%%%%%%%%%%
From Eqs.\ (\ref{grep}) and (\ref{hwave}) follows Faraday's law of induction
%%%%%%%%%%%%%%%%
\begin{equation}
\nabla \times G = n \, H
\label{faraday}
\end{equation}
%%%%%%%%%%%%%%%%
and subsequently the defining wave equation (\ref{green}) of the Green's function. Faraday's law also reveals that $H$ describes the magnetic field strength divided by 
$\mbox{i} c \mu_0 k$ where $c$ denotes the speed of light and $\mu_0$ the permeability of vacuum. Consider a special case of wave propagation first.

{\it Source at origin.---} Imagine the source is placed at the center of the fish eye (and the image (\ref{image}) would be at $\infty$), 
%%%%%%%%%%%%%%%%
\begin{equation}
\mathbf{r_0} = 0 \,,\quad \iota = \delta (\mathbf{r})   \,.
\label{special}
\end{equation}
%%%%%%%%%%%%%%%%
Similar to the electromagnetic wave emitted by a point dipole in free space \cite{Jackson}, we represent $H$ by the ansatz
%%%%%%%%%%%%%%%%
\begin{equation}
H = \nabla \times 2  D (r) \, \mathds{1} \,.
\label{ansatz}
\end{equation}
%%%%%%%%%%%%%%%%
In free space, $2D$ describes the scalar Green's function \cite{Jackson}. In Maxwell's fish eye, $D$ turns out to be the scalar Green's function, as we will show, where the factor of $2$ represents the refractive index (\ref{fish}) at the source. We apply the expression
%%%%%%%%%%%%%%%%
\begin{equation}
\nabla f(r) = \frac{\mathbf{r}}{r}\, \partial_r f
\label{radgrad}
\end{equation}
%%%%%%%%%%%%%%%%
for the gradient of any radial function where $\partial_r$ abbreviates the derivative with respect to the radius, and obtain
%%%%%%%%%%%%%%%%
\begin{equation}
H = \mathbf{r} \times \frac{2\, \partial_r  D}{r} \, \mathds{1} \,.
\label{hr}
\end{equation}
%%%%%%%%%%%%%%%%
We calculate the curl of $H$ by expanding the double vector product in $\nabla \times (\nabla \times 2D \mathds{1})$, using the radial gradient (\ref{radgrad}) and the standard expression of the Euclidean Laplacian in spherical coordinates, 
%%%%%%%%%%
\begin{eqnarray}
\lefteqn{\nabla \times H = \nabla \otimes \frac{\mathbf{r}}{r} \, \partial_r 2D -  \mathds{1} \left(\partial_r^2+\frac{2}{r}\partial_r \right) 2D}
\nonumber\\
& = & \frac{\mathbf{r} \otimes \mathbf{r}}{r^2} \left( \partial_r^2 - \frac{1}{r} \partial_r \right) 2 D - \mathds{1} \left( \partial_r^2 + \frac{1}{r} \partial_r \right) 2 D \,.
\label{curl2}
\end{eqnarray}
%%%%%%%%%%
Then we turn to $\nabla\times n^{-1} \nabla \times H$ that occurs in the wave equation (\ref{hwave}) of the magnetic field. For the first term in expression (\ref{curl2}) the only non--zero contribution to the curl originates from $\nabla \times \mathbf{a} \otimes \mathbf{r} = -\mathbf{a} \times  \nabla \otimes \mathbf{r} = -\mathbf{a} \times  \mathds{1}$. For the curl of the second term divided by $n$ we apply the radial gradient (\ref{radgrad}) as in formula (\ref{hr}). Combining these terms we obtain
%%%%%%%%%%
\begin{eqnarray}
\lefteqn{\nabla \times \frac{1}{n} \nabla \times H  =  \frac{2\mathbf{r}}{r^2} \times  \mathds{1} \left( \frac{2}{n \, r} - \partial_r \frac{1}{n} \partial_r \, r \right) \partial_r   D}    \nonumber\\
 & = &  \frac{2 \, n \, \mathbf{r}}{r} \times \mathds{1} \partial_r \left(\frac{1}{r^2 n^3} \, \partial_r \, r^2 \, n \, \partial_r  D - D \right)
\end{eqnarray}
%%%%%%%%%%
where in the last step we used the explicit formula (\ref{fish}) of the fish eye profile. All expressions in the wave equation (\ref{hwave}), including the curl of the delta functions on the right--hand side, have the same matrix structure as the magnetic field (\ref{hr}). We only need to require
%%%%%%%%%%%%%%%%
\begin{equation}
\frac{1}{r^2 n^3} \, \partial_r \, r^2 \, n \, \partial_r \, D  + (k^2-1) \, D = 
- \frac{\delta (\mathbf{r})}{4 \, n}
\label{radial}
\end{equation}
%%%%%%%%%%%%%%%%
for finding the Green's function $G$ in the special case (\ref{special}). But before we write down the solution of the radially symmetric scalar wave equation (\ref{radial}) we cast it in a geometric form and transform it, for investigating perfect imaging of scalar waves in 3D.

{\it Scalar waves.---} The virtual geometry of the ``fish--eye world'' is characterized by the line element
%%%%%%%%%%%%%%%%
\begin{equation}
{\rm d} s = n (r) {\rm d} l    \,.
\label{line}
\end{equation}
%%%%%%%%%%%%%%%%
In this geometry the Laplacian \cite{LeoPhilbin} appears in spherical coordinates $r$, $\theta$ and $\phi$ as
%%%%%%%%%%%%%%%%
\begin{eqnarray}
\sum_a \nabla_a \nabla^a & = & \frac{1}{r^2 n^3} \, \partial_r \, r^2 \, n \, \partial_r + \frac{1}{r^2} \, \partial_{\theta}^2 + \frac{1}{r^2 \sin^2 \theta} \, \partial_{\phi}^2  \nonumber\\
& = & \frac{1}{n^3} \nabla \cdot n \nabla
\end{eqnarray}
%%%%%%%%%%%%%%%%
in Cartesian coordinates. Consequently, we can write the radial wave equation (\ref{radial}) as
%%%%%%%%%%%%%%%%
\begin{equation}
\left( \frac{1}{n^3} \nabla \cdot n \nabla + k^2-1  \right) D = - \frac{\iota}{n^3}  \,.
\label{scalar}
\end{equation}
%%%%%%%%%%%%%%%%
This is the wave equation of the scalar Green's function. The equation is invariant under coordinate transformations that preserve the geometry, the line element (\ref{line}), because it is entirely composed of geometrical constructions --- the left--hand side contains the Laplacian and $k^2-1$, a scalar,
while the right--hand side consists of delta functions divided by the density $n^3$, the measure of length (\ref{line}) cubed. Such coordinate transformations are the M\"obius transformations
%%%%%%%%%%%%%%%%
\begin{equation}
\mathbf{r'} = \frac{\mathbf{r} (1+r_0^2) - \mathbf{r_0} (1+2 \, \mathbf{r}\cdot\mathbf{r_0} - r^2)}{1+2 \, \mathbf{r}\cdot\mathbf{r_0} + r^2 r_0^2}    \,,
\label{moebius}
\end{equation}
%%%%%%%%%%%%%%%%
because one verifies that
%%%%%%%%%%%%%%%%
\begin{equation}
{\rm d} s' = {\rm d} s    \,.
\end{equation}
%%%%%%%%%%%%%%%%
In 2D, the M\"obius transformations correspond to the rotations on the surface of the 3D sphere in stereographic projection on the plane \cite{Needham}. In 3D, the M\"obius transformations (\ref{moebius}) describe the rotations on the 3D surface of the 4D hypersphere, the virtual geometry of Maxwell's fish eye \cite{Luneburg}. As the relationship (\ref{image}) between source and image is also invariant under M\"obius transformations,
all scalar Green's functions are simply M\"obius--transformed solutions of the radial wave equation (\ref{radial}) with $r$ replaced by the radius
%%%%%%%%%%%%%%%%
\begin{equation}
r' = |\mathbf{r'}| = \frac{|\mathbf{r}-\mathbf{r_0}|}{\sqrt{1+2 \, \mathbf{r}\cdot\mathbf{r_0} + r^2 r_0^2}}   \,.
\label{radius}
\end{equation}
%%%%%%%%%%%%%%%%
A suitable solution is
%%%%%%%%%%%%%%%%
\begin{eqnarray}
D & = & \frac{1}{8\pi} \left( r' + \frac{1}{r'} \right)  \exp (2 \mbox{i} k \arctan r')
\nonumber\\
& = & \frac{1}{8\pi} \left( r' + \frac{1}{r'} \right) \left( \frac{1+\mbox{i}\,r'}{1-\mbox{i}\,r'}  \right)^k  \,,
\label{scalargreen}
\end{eqnarray}
%%%%%%%%%%%%%%%%
a curious variation of the free--space scalar Green's function \cite{Jackson}. The Green's function (\ref{scalargreen}) has two singularities, one at $r'=0$ and the other at $r'= \infty$; one singularity at the source point $\mathbf{r_0}$ where the expression (\ref{radius}) vanishes and the other singularity at the image (\ref{image}) where the transformed radius (\ref{radius}) diverges. The phase $2 k \arctan r'$ is finite, 
because the ``fish--eye world'' is closed --- the surface at the hypersphere is finite. At the image we obtain the phase delay (\ref{phase}). The prefactor of the scalar Green's function (\ref{scalargreen}) was chosen to give delta functions on the right--hand side of the wave equation (\ref{scalar}). The Green's function (\ref{scalargreen}) is analytic in $k$ and decays exponentially on the upper half complex $k$ plane: $D$ is causal. All our requirements are met: scalar waves with the wave equation (\ref{scalar}) and the fish--eye profile (\ref{fish}) support perfect imaging.

{\it Green tensor.---} Having established the scalar Green's function $D$ we follow a similar procedure to deduce the electromagnetic Green tensor $G$.
First, we write down the Green's function for the source at the origin in M\"obius--primed coordinates. One verifies that formula (\ref{grep}) with expression (\ref{curl2}) appears as
%%%%%%%%%%%%%%%%
\begin{equation}
G = \frac{\nabla \times n(r') \nabla \otimes \nabla_0\, D(r') \times \!\stackrel{\longleftarrow}{\nabla_0}}{n(r)\, n(r_0)\, k^2} - \frac{\iota \mathds{1}}{n \, k^2}
\label{result}
\end{equation}
%%%%%%%%%%%%%%%%
evaluated at $r_0=0$. Here $\nabla_0$ denotes the gradient operator with respect to the source point $\mathbf{r}_0$ and the arrow indicates that all terms on the left of it are to be differentiated. Second, all we need to do for establishing the Green's function for an arbitrary source point $\mathbf{r}_0$ is M\"obius--boosting the special case (\ref{result}). We know that $r'$ is M\"obius--invariant, but we also need to transform the bi--tensor components of $G$. For this, we write our result (\ref{result}) in geometric terms for the ``fish--eye world'' with line element (\ref{line}), which is most easily done in index notation. We use the permutation symbol $[abc]$ for the Cartesian curls \cite{LeoPhilbin} and express them by two Levi--Civita tensors $\epsilon^{abc}$ \cite{LeoPhilbin} in the fish--eye geometry, for spectator and source points separately. Finally we lower the first index of each Levi--Civita tensor by the corresponding metric tensors $n(r)^2 \mathds{1}$ and $n(r_0)^2\mathds{1}$, respectively. In formulae,
%%%%%%%%%%%%%%%%
\begin{eqnarray}
G_{ab} & = &
 \sum_{cdef} \frac{[acd]\,[bef]}{n(r)\,n(r_0)\,k^2} \, \frac{\partial^2 n(r')}{\partial x^c\, \partial x^e_0}\, \frac{\partial^2D(r')}{\partial x^d\,\partial x^f_0} - \frac{\iota \delta_{ab}}{n \, k^2}
\nonumber\\
& = &
\sum_{cdef} \frac{\epsilon_a^{\ cd}\,\epsilon_b^{\ ef}}{k^2} \, \frac{\partial^2 n(r')}{\partial x^c\, \partial x^e_0}\, \frac{\partial^2D(r')}{\partial x^d\,\partial x^f_0} - \frac{\iota \delta_{ab}}{n \, k^2} \,.
\label{bitensor}
\end{eqnarray}
%%%%%%%%%%%%%%%%
Formula (\ref{bitensor}) describes a perfect bi--tensor in the fish--eye geometry that remains invariant after M\"obius transformation. Consequently, we can simply drop the qualification $r_0=0$: our result (\ref{result}) is valid for arbitrary source points. Note that the delta currents (\ref{iota}) in expression (\ref{result}) conveniently cancel the delta functions arising in the derivatives of $D$; the Green's function $G$ is singular at source $\mathbf{r}_0$ and image $\mathbf{r}_0'$ but does not develop delta peaks. As the scalar Green's function $D$ is analytic and exponentially decaying on the upper half complex $k$ plane, so is the Green tensor $G$. Source, image (\ref{image}) and phase delay (\ref{phase}) are the same as in the scalar case as well. In short, Maxwell's fish eye in three dimensions perfectly images electromagnetic waves.

{\it Mirror.---} In practice, the fish--eye profile (\ref{fish}) poses a formidable challenge: it is infinitely extended across space and the refractive index $n<1$ for $r>1$, the speed of light exceeds $c$ here and approaches infinity. Both problems are related: Maxwell's fish eye represents a finite virtual space, the 3D surface of the 4D hypersphere, stretched out to infinite physical space. Light can only reach infinity in a finite time with infinite speed. But, one can solve both problems in one stroke by placing a mirror at the unit sphere $(r=1)$ \cite{LeoPerfect}. In this case, the device occupies a finite space, the interior of the unit sphere, and the refractive index ranges from 1 at the mirror to 2 in the center. The same trick has been applied in non--Euclidean cloaking \cite{LeoTyc}. The mirror creates the illusion that light propagates beyond the unit sphere, whereas in reality it is reflected at the mirror. After another reflection, the light focuses at the mirror image of the original focusing point (\ref{image}). 
To show this, we essay employing the inversion in the unit sphere as a mirror transformation of the spectator points,
%%%%%%%%%%%%%%%%
\begin{equation}
\mathbf{r} \rightarrow \frac{\mathbf{r}}{r^2}
\label{inversion}
\end{equation}
%%%%%%%%%%%%%%%%
or, in spherical coordinates, $r \rightarrow r^{-1}$. The mirror--image of the image (\ref{image}) would appear at 
%%%%%%%%%%%%%%%%
\begin{equation}
\mathbf{r_0''} = - \mathbf{r_0}   \,.
\end{equation}
%%%%%%%%%%%%%%%%
The inversion (\ref{inversion}) also preserves the line element (\ref{line}) of Maxwell's fish eye and hence the transformed scalar Green's function remains a solution of the wave equation (\ref{scalar}). We only need to transform the spectator indices of the Green tensor by the matrix $P$ (but not the source indices). As reflection gives rise to a phase shift of $\pi$ \cite{Jackson} we substract the reflected field from the original one, 
%%%%%%%%%%%%%%%%
\begin{equation}
G' = G(r) - P \, G(r^{-1})     \,.
\label{mirrorgreen}
\end{equation}
%%%%%%%%%%%%%%%%
In Cartesian coordinates,
%%%%%%%%%%%%%%%%
\begin{equation}
P = r^2 \, \mathds{1} - 2 \, \mathbf{r} \otimes \mathbf{r}
\end{equation}
%%%%%%%%%%%%%%%%
but in spherical coordinates
%%%%%%%%%%%%%%%%
\begin{equation}
P^{i}_{\,\,j} = {\rm diag} \left( -r^2, 1, 1  \right)  \,.
\end{equation}
%%%%%%%%%%%%%%%%
Consequently, at the unit sphere the tangential components of the Green's function, representing the electric field, vanish, which is the defining property of a perfect mirror. The Green's function (\ref{mirrorgreen}) obeys not only the wave equation (\ref{green}) but also the correct boundary conditions at the spherical mirror, which justifies formula (\ref{mirrorgreen}). 

Finally, we may also include absorption in our theory, although in a simplified model. Absorption is described by the imaginary part $n''$ of the refractive index. Suppose $n''(r)$ is proportional to the real part given by the fish-eye profile (\ref{fish}). This case is equivalent to having a complex wavenumber $k$ in the definition (\ref{green}) of the Green's function. The singularities of the Green's function (\ref{scalargreen}) describe source and image, but they are not affected by the wavenumber that only reduces the amplitude: the imaging quality is resistant to absorption, at least in our simple model. The fish--eye mirror in 3D images with a resolution no longer limited by the wave nature of light.

{\it Acknowledgements.---} We are indebted to Lucas Gabrielli, Michal Lipson and Tom\'a\v{s} Tyc for inspiring discussions. Our work is supported by the Royal Society of Edinburgh, the Scottish government, a Wolfson Award and a Theo Murphy Blue Skies Award of the Royal Society of London.

%%%%%%%%%%%%%%%%%%%%%%%%%%%%%%%%%%%%%%%%%%%%%%%%%%%%%%%%%%%%%%%%%%%%%%%%%%%%%%%%%%%
\end{document}